\documentclass[twocolumn]{aastex63}

\usepackage{amsmath}
\usepackage{amssymb}
\usepackage{natbib}
\usepackage{graphicx}
\usepackage{color}
\usepackage{hyperref}
\usepackage{booktabs}

\usepackage{savesym}
\savesymbol{tablenum}
\usepackage{siunitx}
\restoresymbol{SIX}{tablenum}

\DeclareSIUnit{\erg}{erg}

\newcommand*{\ra}[2][]{{
    \def\SIUnitSymbolDegree{\textsuperscript{h}}%
    \def\SIUnitSymbolArcminute{\textsuperscript{m}}%
    \def\SIUnitSymbolArcsecond{\textsuperscript{s}}%
    \ang[#1]{#2}}%
}

\begin{document}

\title{Detection of Galactic and Extragalactic Millimeter-Wavelength Transient Sources with SPT-3G}

\author{S.~Guns}
\affiliation{Department of Physics, University of California, Berkeley, CA, 94720, USA}
\author{A.~Foster}
\affiliation{Department of Physics, Case Western Reserve University, Cleveland, OH, 44106, USA}
\author{C.~Daley}
\affiliation{Department of Astronomy, University of Illinois at Urbana-Champaign, 1002 West Green Street, Urbana, IL, 61801, USA}
\author{A.~Rahlin}
\affiliation{Fermi National Accelerator Laboratory, MS209, P.O. Box 500, Batavia, IL, 60510, USA}
\affiliation{Kavli Institute for Cosmological Physics, University of Chicago, 5640 South Ellis Avenue, Chicago, IL, 60637, USA}
\author{N.~Whitehorn}
\affiliation{Department of Physics and Astronomy, Michigan State University, East Lansing, MI 48824, USA}
\affiliation{Department of Physics and Astronomy, University of California, Los Angeles, CA, 90095, USA}
\author{P.~A.~R.~Ade}
\affiliation{School of Physics and Astronomy, Cardiff University, Cardiff CF24 3YB, United Kingdom}
\author{Z.~Ahmed}
\affiliation{Kavli Institute for Particle Astrophysics and Cosmology, Stanford University, 452 Lomita Mall, Stanford, CA, 94305, USA}
\affiliation{SLAC National Accelerator Laboratory, 2575 Sand Hill Road, Menlo Park, CA, 94025, USA}
\author{E.~Anderes}
\affiliation{Department of Statistics, University of California, One Shields Avenue, Davis, CA 95616, USA}
\author{A.~J.~Anderson}
\affiliation{Fermi National Accelerator Laboratory, MS209, P.O. Box 500, Batavia, IL, 60510, USA}
\affiliation{Kavli Institute for Cosmological Physics, University of Chicago, 5640 South Ellis Avenue, Chicago, IL, 60637, USA}
\author{M.~Archipley}
\affiliation{Department of Astronomy, University of Illinois at Urbana-Champaign, 1002 West Green Street, Urbana, IL, 61801, USA}
% ...


\begin{abstract}
    High-angular-resolution cosmic microwave background experiments provide a unique opportunity to conduct a survey of time-variable sources at millimeter wavelengths, a population which has primarily been understood through follow-up measurements of detections in other bands. Here we report the first results of an astronomical transient survey with the South Pole Telescope (SPT) using the SPT-3G camera to observe 1500 square degrees of the southern sky. The observations took place from March to November 2020 in three bands centered at \SIlist{95;150;220}{GHz}. This survey yielded the detection of fifteen transient events from sources not previously detected by the SPT. The majority are associated with variable stars of different types, expanding the number of such detected flares by more than a factor of two. The stellar flares are unpolarized and bright, in some cases exceeding 1 Jy, and have durations from a few minutes to several hours. Another population of detected events last for 2--3 weeks and appear to be extragalactic in origin. Though data availability at other wavelengths is limited, we find evidence for concurrent optical activity for two of the stellar flares. Future data from SPT-3G and forthcoming instruments will provide real-time detection of millimeter-wave transients on timescales of minutes to months.
\\
\end{abstract}

\section{Introduction}

\label{sec:intro}

Long-wavelength (infrared and longer) transient sources have been predicted to be a powerful source of information on a wide class of high-energy astrophysical objects, including gamma-ray burst afterglows, the jet launch area of active galactic nuclei (AGN), tidal-disruption events, stellar flares, and more \citep[e.g.][]{metzger15}.
Ongoing efforts to deploy dedicated transient surveys at longer-than-optical wavelengths have already yielded detections of galactic and extragalactic transients in the near infrared \citep{de20} and at radio frequencies \citep{mooley16,law18,lacy20}.

The transient millimeter-wavelength (mm-wave) sky is currently largely unexplored except in follow-up observations of sources detected first at other wavelengths \citep{metzger15}. 
Telescopes designed for observations of the cosmic microwave background (CMB) are optimized for wide-field surveys, operate at millimeter wavelengths, and have a typical observing strategy in which a patch of sky up to thousands of square degrees is re-observed regularly, making them powerful instruments for transient surveys \citep{holder19}.

\citet{whitehorn2016} performed the first transient survey with a CMB experiment using SPTpol, the second-generation camera on the South Pole Telescope (SPT),
and found one transient candidate with no apparent counterpart.
Recently, the Atacama Cosmology Telescope (ACT) serendipitously discovered three flares associated with stars \citep{naess2021} similar to those previously observed by \cite{brown&brown06,beasley&bastian98,massi06,salter10,bower03}.

In this paper we report the first results from a transient-detection program using SPT-3G, the third-generation camera on the SPT, during the austral winter of 2020.
This analysis improves on the sensitivity of the earlier SPTpol study through substantial improvements in observing cadence, survey area, wavelength coverage, and point-source sensitivity.
In this study, we found 15 unique transient events: 13 short-duration events associated with an eclectic mix of 8 nearby stars (with 3 stars having multiple events) and 2 longer-duration events of likely extragalactic origin. The transient events associated with stars have very large ($>100\times$) increases in mm-wave luminosity over the source's quiescent state, with peak flux densities exceeding \SI{1}{Jy} in some cases, placing them among the brightest mm-wave objects in the SPT-3G footprint when they are flaring.

\section{The SPT Instrument and Survey}
\label{sec:survey}

The SPT is a 10-meter telescope located at Amundsen-Scott South Pole Station, Antarctica, and is optimized to survey the CMB at mm wavelengths \citep{carlstrom11}.
The SPT-3G camera consists of \num{\sim 16000} multichroic, polarization-sensitive bolometric detectors which operate in three bands across the atmospheric transmission windows at \SIlist{95;150;220}{GHz}, with angular resolution of $\sim$1 arcminute.
The SPT-3G survey covers a \SI{1500}{\deg^2} footprint spanning \SIrange{-42}{-70}{\degree} in declination (Dec) and \SIrange{-50}{50}{\degree} in right ascension (RA), and has been observed in the current configuration since 2018 \citep{dutcher18}.
We observe this footprint on a cadence set by the 16-hour observing day (limited by the cryogenic refrigeration cycle). 

Due to detector responsivity and linearity constraints from atmospheric loading, the SPT-3G footprint is broken up into 4 subfields centered at declinations of $-\ang{44.75}, -\ang{52.25}, -\ang{59.75}$, and $-\ang{67.25}$.
Each subfield is observed by rastering the telescope in scans at constant elevation, taking an \ang{;11.25;} step in elevation and repeating until the full elevation range of the subfield has been observed.
This process takes approximately {2} {hours}.
During an observing day, two subfields are observed three times each, with the remaining time in the observing day used for calibration observations and detector re-tuning.
As a result, the re-observation cadence of a given point in the field ranges from {2} {hours} to {20} {hours}. 
Our cadence is chosen to reach a uniform survey depth between each of the subfields over the course of an observing season. While the cadence is not optimally designed for transient searches, it still allows us to have rapid (${\sim} 2$-hour) near-daily observations over a 36-week observing season and to effectively probe flaring sources at a variety of timescales. For more detailed information on the SPT-3G experiment, see \cite{bender18}.

\section{Methods}
\label{sec:methods}
\subsection{Transient Detection}
\label{sec:detection}

To detect transient sources, we construct maps of each subfield observation using a pipeline similar to the one used for analysis of the CMB power spectrum \citep{dutcher21}. 
This pipeline applies a number of filtering steps to reduce low-frequency noise (primarily from atmospheric emission) in the detectors' time-ordered data (TOD), then weights and bins the TOD into an intensity map of the field.

After the map binning step, we make difference maps by subtracting a year-long average map of the survey field, constructed using 2019 SPT-3G data. 
This effectively removes all static backgrounds including, but not limited to, the CMB, galaxy clusters, and non-time-varying point sources. 
Many of the AGN in the footprint are variable. To prevent detections of variability in bright AGN, we mask all point sources that had an average flux above \SI{5}{mJy} in 2019 SPT-3G data, using a mask radius of up to \ang{;5;}, depending on brightness.
We apply an additional noise filter using a weighted convolution in map space. The filter uses an annulus with an outer radius of \ang{;5;} and an inner radius of \ang{;2;}, and acts as a high-pass filter to remove noise on scales larger than \ang{;5;} without subtracting signal power contained within \ang{;2;} of each map location.
Finally, we apply another real-space convolution filter using a beam template to maximize sensitivity to point sources. 
The beam template is constructed from measurements of in-field bright point sources and dedicated Saturn observations in a manner similar to \citet{dutcher21}.

For a given location (map pixel) on the sky, we consider its multi-band flux density as a function of time $\phi^b_t$ ($b \in \{ \SI{95}{GHz}, \SI{150}{GHz} \}$). We then use a multi-band extension of the maximum-likelihood transient finding method used in our previous study \citep{whitehorn2016} and derived from that used in \cite{braun10}. The \SI{220}{GHz} maps have median noise levels that are on average five times higher than the other bands, and for reasonable flare spectra make negligible contributions to the total sensitivity. To save on processing time they are excluded from the likelihood, which is especially important for any live search for which local computing resources at the South Pole are limited. We inspect the \SI{220}{GHz} data for a given flare post-detection only. 
The multi-band likelihood takes the form

\begin{equation}
    -2\,\mathrm{ln}\mathcal{L}(f) = \sum_t \sum_b \left[ \frac{\phi_t^b - f_t^b}{\sigma_t^b} \right ]^2, 
\label{eq:llh}
\end{equation}
where $f$ is the time-domain flare model and $\sigma^b_t$ is the map noise estimate for the given band, pixel, and time. We use a Gaussian ansatz for the flare model $f$ with independent amplitudes for each band \citep{whitehorn2016} to provide a smoothly optimizable function for the detection---but not parameter estimation---of flaring sources:

\begin{equation}
f^b_t \equiv f(t; S_b, t_0, w) = S_b \exp \left[-\frac{(t-t_0)^2}{2w^2}\right], 
\label{eq:llhgauss}
\end{equation}
where $S_b$ is a flux density (in mJy) for each band $b$, $t_0$ is the event time and $w$ the flare width. The test statistic that is used to infer significance is the ratio of the likelihood function at the extremal (best-fit) parameter values to the null hypothesis likelihood at zero amplitude, with an additional term to account for the statistical preference
for short duration events:

\begin{equation}
    \mathrm{TS} \equiv 2\,\mathrm{ln}\mathcal{L}(f) - 2\,\mathrm{ln}\mathcal{L}(0) + 2\,\mathrm{ln}\Bigl(\frac{w}{\Delta T}\Bigr )
\label{eq:teststat}
\end{equation}
where $\Delta T$ is the total duration of the data set and $\mathcal{L}(0)$ is the likelihood at zero amplitude for both bands, which does not depend on either $t_0$ or $w$. The third term (the width penalty) is necessary because, as the flare width $w$ gets smaller, there are many more uncorrelated starting times $t_0$, leading to a maximization bias in the likelihood for short flare widths.
To remove the bias, we apply a likelihood penalty term $P(w) \sim \mathrm{ln}(w)$ which is approximately equivalent to marginalizing over a uniform prior in $w$ \citep{braun10}.

We maximize Equation \eqref{eq:teststat} to find the best-fit parameters $(\hat{S}_b,\hat{t}_0,\hat{w})$ for a candidate event.
Following Wilks' theorem, the TS value is approximately $\chi^2$-distributed, with a number of degrees of freedom obtained by fitting to the distribution of negative fluctuations in the maps, which are signal-free and match the distribution of positive fluctuations in the noise-dominated low-TS region.
We then place a cut on TS and report here all events with $\mathrm{TS} > 100$, corresponding to a $9.7 \sigma$ detection (see Section \ref{sec:backgrounds} for more discussion).

\begin{figure*}
\plotone{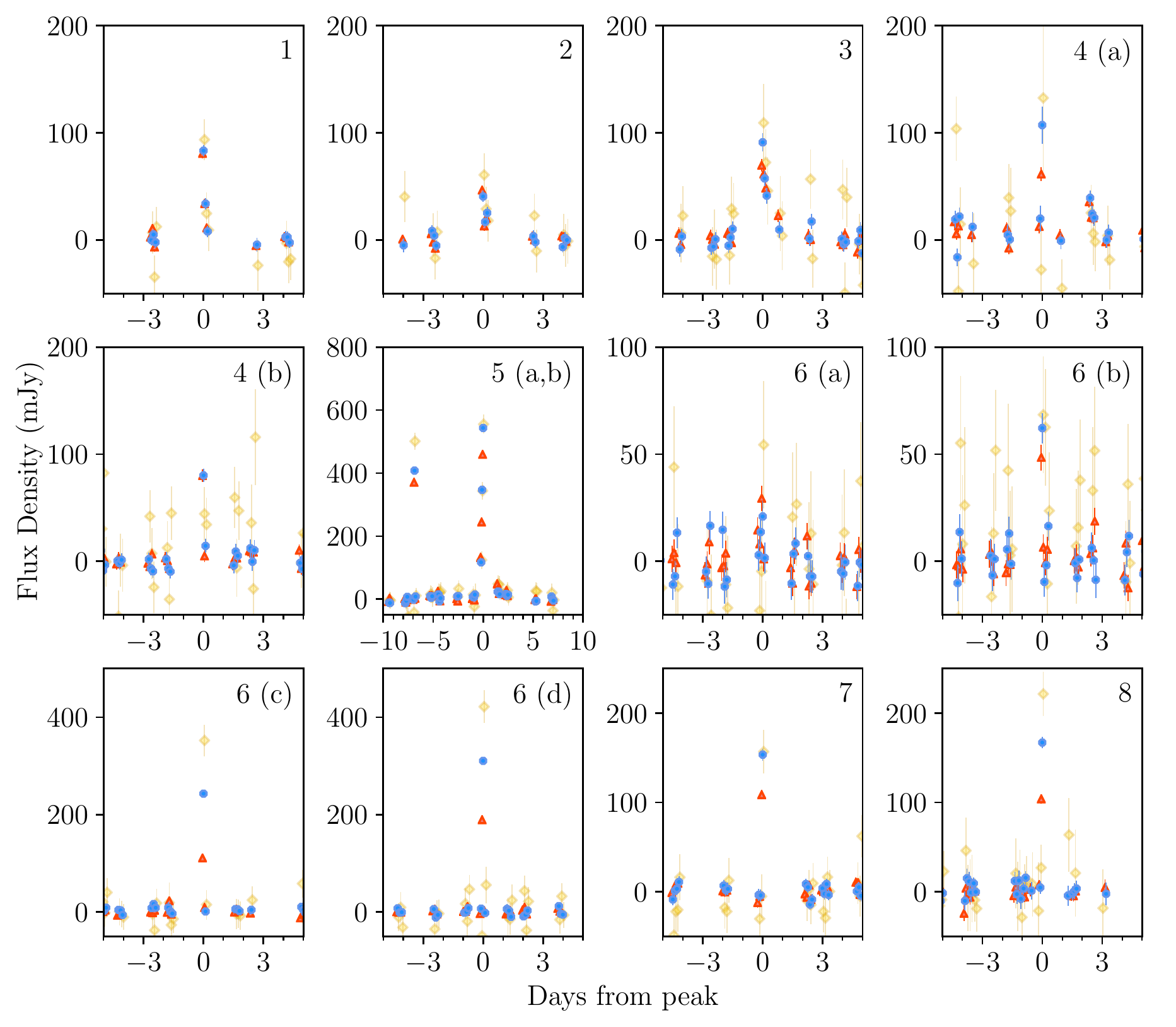}

\caption{
  Lightcurves of the 13 stellar flares observed in this study at \SI{95}{GHz} (\textit{red}), \SI{150}{GHz} (\textit{blue}), and \SI{220}{GHz} (\textit{gold}). 
  Two of the flares for Z~Ind were close in time and are shown together in one panel. 
  In all cases, the rise and fall times are short (hours or less) and the spectra approximately flat.}
\label{fig:allflares}
\end{figure*}

Using computing resources at the South Pole, \ang{;0.25;}-resolution maps are automatically created and filtered following every two-hour subfield observation. 
We then construct lightcurves stretching back \SI{14}{days} for each pixel in the subfield. Due to the observing cadence outlined in Section~\ref{sec:survey}, lightcurves tend to have 2-3 clustered data points followed by a gap in time, giving rise to the characteristic time coverage seen in Figure~\ref{fig:allflares}.
We run the flare fitting algorithm  on each lightcurve and flag significant events for further analysis.  
This analysis pipeline allows for flare detection within no more than twelve hours from the flare time, giving us the ability to send out online alerts to recommend follow-up in other bands. 
The majority of sources described in this article were observed before this online detection system was activated, but were analyzed using the same pipeline after the fact.

When a transient event is detected, we generate polarized flux density lightcurves in order to determine the peak event amplitude, spectral index, and polarization fraction.
To estimate the spectral index $\alpha$, we fit a frequency-dependent flux model $\phi^b = \phi^{\SI{150}{GHz}} (\frac{b}{\SI{150}{GHz}})^\alpha$ to the three bands using a $\chi^2$ metric.
We marginalize over the \SI{150}{GHz} flux density by using the best-fit $\phi^{\SI{150}{GHz}}$ for each $\alpha$, and estimate $1\sigma$ confidence intervals using a $\Delta \chi^2 = 1$ criterion.
Polarization fractions are estimated from the Stokes-$Q$ and Stokes-$U$ lightcurves using the maximum-likelihood approach of \citet{vaillancourt06} to reduce noise bias on the polarized flux density $P = \sqrt{Q^2 + U^2}$.

\subsection{Backgrounds}
\label{sec:backgrounds}

The sources presented in this paper belong to a small category of emitters that show highly significant events ($\mathrm{TS} > 100$, corresponding to $>9.7 \sigma$) and are, with two exceptions, detected in multiple observations, ruling out an origin in an instrumental glitch, satellite pass, or other noise source. 
At this level of significance, the expected rate of events from Gaussian fluctuations sourced by instrumental and atmospheric noise is below 1 event per million years of observing.
False detections may also be caused by non-astrophysical in-band emitters; these are expected to be the dominant noise source at short timescales. 
The only such contamination we unambiguously detected at this high significance level was thermal emission from weather balloons launched from the South Pole station by the Antarctic Meteorological Research Center (AMRC)\footnote{\url{https://amrc.ssec.wisc.edu}} and the National Oceanic and Atmospheric Administration (NOAA).\footnote{\url{https://www.esrl.noaa.gov/gmd/ozwv/ozsondes/spo.html}} A small fraction of these balloons drifted through the telescope's field of view shortly after launch. 
Based on the typical launch cadence and our observing strategy we expect to detect $\mathcal{O}(10)$ such balloons over the course of an observing season. 
Weather balloons can show up in maps at brightnesses exceeding \SI{1}{Jy}. 
Typically, their proximity and fast movement creates recognizable extended structure over scales of many arcminutes. 
Rather than implement a cut on this signal within an observation, we placed a requirement as part of our detection pipeline that sources appear at a signal-to-noise ratio $> 3$ and a fixed location in more than one independent observation. This makes the search presented here insensitive to contamination by man-made sources which are not fixed in RA/Dec and rapidly move out of the field, at the cost of reducing sensitivity to rapid events. 
Two of the 15 detected flares (source 7 and 8) were originally detected by other means, as part of debugging the analysis pipeline, and are included here despite being detected in only one observation. 
In the case of these two sources, a number of other checks were performed (looking at sub-observation-scale data to ensure the source was stationary for a significant amount of time and cross-checking with balloon launch schedules) to ensure the astrophysical origin of the flare. 
In the future, we expect to be able to automate both weather balloon detection and sub-observation timescale analysis, and relax the multiple-observation requirement across the board.

\section{Results}
\label{sec:results}

\begin{table*}
\sisetup{table-align-uncertainty, minimum-integer-digits = 2} 
\centering

\begin{tabular*}{\textwidth}{l @{\extracolsep{\fill}} cclSSSr}
\toprule
\multicolumn{4}{c}{} & \multicolumn{3}{c}{Peak Flux Density (mJy)} & \multicolumn{1}{c}{}\\
\cmidrule{5-7}

           ID &  RA  & Dec & Time (UTC) 
           & \SI{95}{GHz}& \SI{150}{GHz} & \SI{220}{GHz} 
           & {$\mathrm{TS}$} \\
\midrule

1          & \ra{23;20;47.6} & \ang{-67;23;23} & 2020-03-26 02:25 
           & 81 \pm 4 & 83\pm5  & 93\pm19 
           & 709 \\ \addlinespace

2          & \ra{23;13;53.1} & \ang{-68;17;34} & 2020-04-01 18:12             
           & 46\pm4 & 40\pm5  & 61\pm20 
           & 213 \\ \addlinespace

3          & \ra{21;01;21.2} & \ang{-49;33;15} & 2020-04-02 14:50 
           & 70\pm6 & 91\pm9  & 103\pm36 
           & 541 \\ \addlinespace

4 (a)      & \ra{21;20;44.5} & \ang{-54;37;56} & 2020-06-03 02:35              
           & 61\pm6 & 108\pm17 & 230\pm69 
           & \\ \addlinespace[-2pt]
\hfill (b) & & & 2020-09-04 09:15 
           & 80\pm6 & 80\pm6 & 44\pm25 
           & 134\\ \addlinespace

5 (a)      & \ra{21;54;23.8} & \ang{-49;56;36} & 2020-06-17 09:20 
           & 370 \pm 6  & 408 \pm 7 & 501 \pm 28 
           & \\ \addlinespace[-2pt]
\hfill (b) & & & 2020-06-24 06:09 
           & 459\pm6  & 543\pm8 & 558\pm30 
           & \num{16416} \\ \addlinespace

6 (a)      &  \ra{02;34;22.4} & \ang{-43;47;53} & 2020-06-21 10:42 
           & 29 \pm 6 & 21 \pm 7 & 54 \pm 30 
           & \\ \addlinespace[-2pt]
\hfill (b) & & & 2020-07-10 08:24 
           & 48 \pm 6  & 62 \pm 7  & 68 \pm 27   
           & \\ \addlinespace[-2pt]
\hfill (c) & & & 2020-09-17 04:51 
           & 111\pm6 & 243\pm8 & 352\pm32  
           & \\ \addlinespace[-2pt]
\hfill (d) & & & 2020-11-05 15:34 
           & 189\pm6 & 310\pm8 & 422\pm33  
           & 2715 \\ \addlinespace

7          &  \ra{02;55;31.6} & \ang{-57;02;54} & 2020-09-18 06:34 
           & 109\pm5 & 154\pm7 & 157\pm25
           & 1119 \\ \addlinespace

8          &  \ra{00;21;28.7} & \ang{-63;51;10} & 2020-11-14 01:47 
           & 104\pm5 & 167\pm7 & 221\pm25 
           & 1184 \\ \addlinespace

\midrule
9          & \ra{22;41;16.7} & \ang{-54;01;07} & 2020-07-08 
           & 13 \pm 2 & 13 \pm 2 &14 \pm 7
           & 221 \\ \addlinespace 

10         & \ra{03;01;16.1} & \ang{-57;19;21} & 2020-07-08 
           & 24\pm2 & 36\pm3 & 40\pm10 
           & 1090 \\ 
\bottomrule
\end{tabular*}
\caption{
  \sisetup{minimum-integer-digits = 1} 
  Transient events detected by SPT-3G between March 23, 2020 and November 15, 2020.
  Each unique source was given a numbered ID, and each flare was labeled by a letter in the case of multiple flares. 
  Source RA and Dec are the best-fit locations measured by SPT-3G.
  The horizontal line differentiates the stellar flares (above) from the long-duration, likely extragalactic transients (below).  
  All sources listed have average flux densities below \SI{5}{mJy} at \SI{150}{GHz} in 2019 SPT-3G data.
  Peak flux densities are averaged over subfield observations and quoted relative to the 2019 average.
  Peak flare times correspond to the beginning of the subfield observation in the case of stellar flares, and to the center of a week-long integration in the case of the long-duration transients.
  The test statistic ($\mathrm{TS}$) value is computed on the full 2020 lightcurve for each source and is shown only for the flare that maximizes the $\mathrm{TS}$ (generally the brightest one); the cut value used in this search is $TS > 100$.
  Several stars showed other flares that had a signal-to-noise $>5$ in at least one observing band and are also shown in this table.
}
\label{tab:events}
\end{table*}

During 3500 hours of observations taken over an 8-month period from 23~March to 15~November, 2020, we observed 10 unique sources with at least one $\mathrm{TS} > 100$ event at locations not associated with point sources previously detected by any SPT survey. This was enforced for SPT-3G by masking point sources with an average \SI{150}{GHz} flux density greater than \SI{5}{mJy} in 2019.

After detecting a source with at least one flare above the $\mathrm{TS}$ threshold, we inspect the lightcurve and tag other flares with signal-to-noise $>5$ in at least one of the observing bands, bringing the total up to the 15 events shown in Table~\ref{tab:events}. The detected flares have emission timescales ranging from tens of minutes to three weeks, with peak brightnesses (averaged over the $\sim$20 minutes of on-source time during the subfield observation) at \SI{150}{GHz} from \SIrange{15}{540}{mJy}, nearing the brightest mm-wave sources in the SPT-3G footprint. 
Given the upper limit on quiescent flux density in 2019 SPT-3G data of \SI{<5}{mJy} for these objects, this represents factors of at least \numrange{4}{100} increase in luminosity above the sources' quiescent states. 

\begin{figure}
\includegraphics[clip, trim = 0 -0.6cm 0 -0.6cm, width=\linewidth]{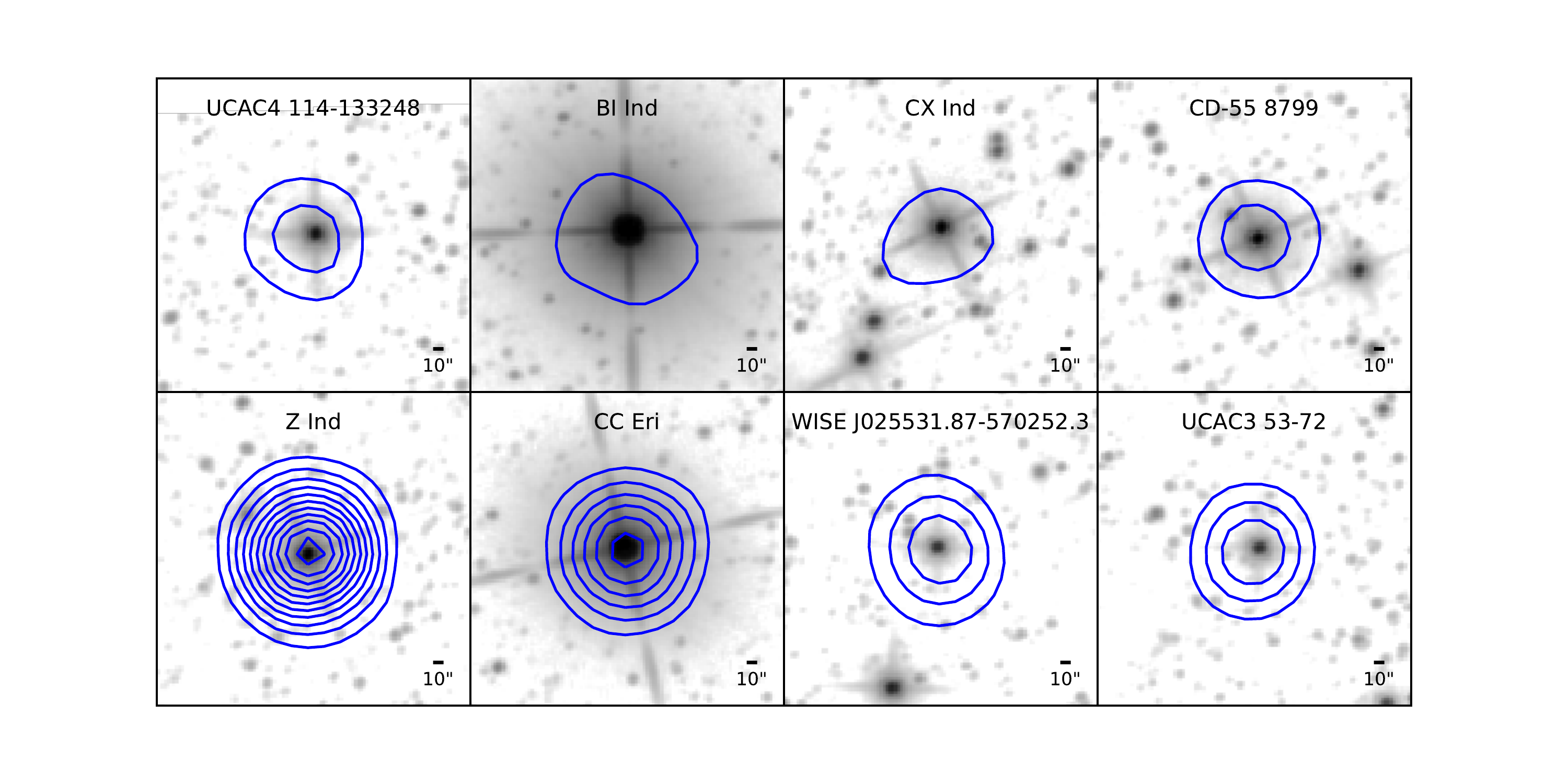}
\caption{Grayscale images of associated stars from unWISE \SI{3.4}{\mu m} W1 \citep{Lang_2014} in log stretch. Blue contours show the SPT-3G \SI{150}{GHz} flux density contours in steps of 5$\sigma$ from the peak signal. The extended cross-like features are diffraction spikes. }
\label{fig:wise_panel_stars}
\vspace*{2cm}
\end{figure}

The detected objects are split into two classes. 
The majority (13 flares from 8 objects) are associated with stars of a wide variety of types (Section~\ref{sec:stars}) and reminiscent of a small number of reports in the literature by e.g. \cite{naess2021,massi06,brown&brown06} of serendipitously detected mm-wave stellar flares. 
Stellar flare associations in WISE with SPT-3G flux density contours are shown in Figure~\ref{fig:wise_panel_stars}.
The fast timescale of emission (from tens of minutes to hours) and approximately flat spectra are suggestive of synchrotron emission, but the lack of detectable linear polarization and sometimes-rising spectra (Table~\ref{tab:specind_polfrac}) imply the emission region is likely inhomogeneous or optically thick for at least part of the observing period.
The remaining two events are not spatially coincident with any cataloged galactic sources, suggesting an extragalactic origin. 
These two sources (Section~\ref{sec:extragalactic}) had triangular light curves lasting \SIrange{2}{3}{weeks}, with flat spectra and peak flux densities between \SIrange{15}{40}{mJy}.
Due to the high instantaneous signal-to-noise on the short-duration flares, the typical positional uncertainties for these events are $\lesssim$\SI{10}{arcsec}, leading to unambiguous associations with known variable stars. Position determination for the two long-duration sources is discussed in more detail in Section~\ref{sec:extragalactic}.

\begin{table}
\centering
\begin{tabular*}{\columnwidth}{l @{\extracolsep{\fill}} S[table-format=1.2(3)]SS}
\toprule
\multicolumn{1}{c}{} & \multicolumn{1}{c}{} & \multicolumn{2}{c}{Pol. Frac.  (95\% UL)} \\
\cmidrule{3-4}
ID & {Spectral Index} & \SI{95}{GHz} & \SI{150}{GHz} \\
\midrule
1          & 0.10 \pm 0.16 & < 0.24 & <0.20 \\ \addlinespace
2          &  -0.1 \pm 0.3 & < 0.44 & <0.37 \\ \addlinespace
3          & 0.6 \pm 0.2   & < 0.33 & <0.30 \\ \addlinespace
4 (a)      & 1.0\pm 0.4    & < 0.34 & <0.55 \\ \addlinespace[-2pt]
\hfill (b) & -0.2  \pm 0.2 & < 0.33 & <0.25 \\ \addlinespace
5 (a)      & 0.25 \pm 0.05 & < 0.06 & <0.06 \\ \addlinespace[-2pt]
\hfill (b) & 0.30 \pm 0.04 & < 0.04 & <0.06 \\ \addlinespace
6 (a)      & -0.4 \pm 0.9  & < 0.79 & <1    \\ \addlinespace[-2pt]
\hfill (b) & 0.6  \pm 0.3  & < 0.47 & <0.33 \\ \addlinespace[-2pt]
\hfill (c) & 1.52 \pm 0.10 & < 0.16 & <0.11 \\ \addlinespace[-2pt]
\hfill (d) & 1.04 \pm 0.07 & < 0.12 & <0.09 \\ \addlinespace
7          & 0.71 \pm 0.11 & < 0.16 & <0.14 \\ \addlinespace
8          & 1.07 \pm 0.10 & < 0.31 & <0.14 \\ \addlinespace
\midrule
9          & {Figure~\ref{fig:ldt_specinds}} & <0.46 & <0.59 \\ \addlinespace
10         & {Figure~\ref{fig:ldt_specinds}} & <0.31 & <0.27 \\
\bottomrule
\end{tabular*}
\caption{
  Spectral indices and \SI{95}{\percent} upper-limit polarization fractions for the transient events listed in Table~\ref{tab:events}.
  No sources had statistically significant detections of polarization.
  Polarization fractions shown are calculated only at flare peaks.
  \SI{220}{GHz} polarization fractions are omitted from this table due to low signal-to-noise.
}
\label{tab:specind_polfrac}
\end{table}

\begin{figure}
\centering
\includegraphics[clip, trim = 0 3mm 0 2mm, width=\linewidth]{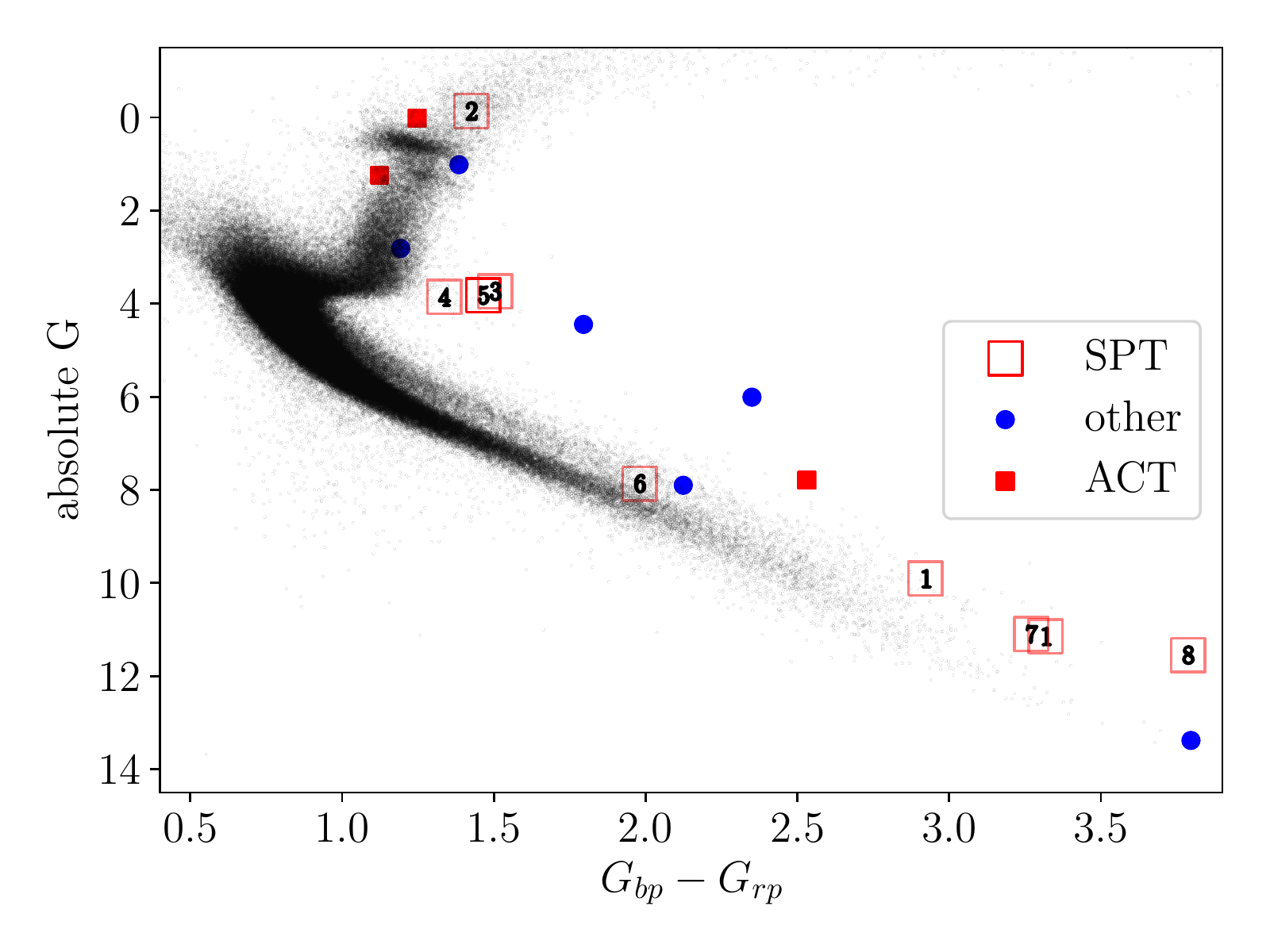}
\caption{Color-magnitude diagram for Gaia stars brighter than apparent magnitude $G=15$ in the SPT-3G footprint.
\textit{Red} symbols show the stars associated with stellar flares detected in CMB surveys; UCAC4 114-133248 (associated with Flare 1) is a double star with both stars measured by Gaia.
\textit{Blue} circles show the previously reported mm-wave stellar flares referenced in Section~\ref{sec:results} that have Gaia counterparts.
\label{fig:hr}}
\end{figure}

A large fraction of the flares were in locations covered by the All-Sky Automated Survey for SuperNovae \citep[ASAS-SN,][]{shappee14,kochanek17}, which provides optical lightcurves with a daily or near-daily cadence. 
For two of the stellar flares, we found evidence for optical activity in the ASAS-SN data. Additionally, one of those two was under observation by the Transiting Exoplanet Survey Satellite \citep[TESS,][]{ricker15} at the time of the flare, providing simultaneous mm-wave and optical coverage of an energetic stellar flare with high time resolution.
The remaining flares were not associated with optical excesses, and none of the events reported here were coincident with reports to alert systems (GCN\footnote{\url{https://gcn.gsfc.nasa.gov/}} or ATel\footnote{\url{http://www.astronomerstelegram.org}}).

\subsection{Stellar flares}
\label{sec:stars}

The observed flares arise in a wide variety of stars. 
The Hertzsprung-Russell (HR) diagram for these stars is shown in Figure~\ref{fig:hr}, using data from the {\em Gaia} mission \citep{gaia_16, gaiadr2_18}. 
Most of the associated stars are known to be X-ray emitters, with counterparts in the ROSAT All-Sky Survey 2RXS catalog \citep{2rxs2016}. 
Only the two M dwarfs, UCAC3~53-724 and WISE~J025531.87-570252.3, are not known X-ray emitters.
By selecting on mm-wave flaring in stars, we appear to be highly biased toward stars that are X-ray sources.
Coronal activity is related to both flaring and X-ray emission so the correspondence is not unexpected.
We randomly selected stars in the SPT-3G footprint with {\em Gaia} apparent magnitude G$<$15 and found that less than $1\%$ had a 2RXS source within \ang{;1;}.
We also found that the probability of a random point being within \ang{;;15} (the furthest SPT-3G flare position association is \ang{;;13}) of a {\em Gaia} source of G$<$15 within the SPT-3G \SI{1500}{\deg^2} footprint is $2\times10^{-3}$, lending further confidence to the associations.

\begin{table*}
\centering
\sisetup{table-format=1.1e2, table-align-exponent=true}
\begin{tabular*}{\textwidth}{l @{\extracolsep{\fill}} lS[table-format=3.3(3)] SSSr}
\toprule
ID & Association      & {Distance (pc)}     & {$\nu L_\nu^{95}$(\si{\erg \per \s})} & {$\nu L_\nu^{150}$(\si{\erg \per \s})} & {$\nu L_\nu^{220}$(\si{\erg \per \s})} &  Type  \\
\midrule

1          & UCAC4 114-133248 & 41.0 \pm 0.1 
           & 1.6e28 & 2.5e28 & 4.1e28 
           & Double M Dwarfs$^*$ \\ \addlinespace

2          & BI Ind & 312 \pm 3 
           & 5.1e29 & 7.0e29 & 1.6e30 
           & RS CVn$^*$ \\ \addlinespace

3          & CX Ind & 235 \pm 3 
           & 4.4e29 & 9.0e29 & 1.5e30 
           & BY Dra Variable$^*$  \\ \addlinespace

4 (a)      & CD-55 8799 & 201 \pm 2 
           & 2.8e29 & 7.9e29 & 2.5e30
           & Rotational Variable$^*$ \\ \addlinespace[-2pt]  
\hfill (b) &  & 
           & 3.7e29 & 5.8e29 & 4.7e29 
           & ~ \\ \addlinespace

5 (a)      & Z Ind & 199 \pm 1    
           & 1.7e30 & 2.9e30 & 5.2e30 
           & Rotational Variable$^*$ \\ \addlinespace[-2pt] 
\hfill (b) & &  
           & 2.1e30 & 3.9e30 & 5.8e30 
           & ~ \\ \addlinespace

6 (a)      & CC Eri & 11.537\pm 0.005 
           & 4.4e26 & 5.0e26 & 1.9e27 
           & BY Dra Variable$^*$\\ \addlinespace[-2pt] 
\hfill (b) & &  
           & 7.3e26 & 1.5e27 & 2.4e27 
           & \\ \addlinespace[-2pt] 
\hfill (c) & &  
           & 1.7e27 & 5.8e27 & 1.2e28 
           & \\ \addlinespace[-2pt] 
\hfill (d) & &  
           & 2.9e27 & 7.4e27 & 1.5e28 
           & \\ \addlinespace

7          & WISE J025531.87-570252.3 & 45.6 \pm 0.1 
           & 2.6e28 & 5.8e28 & 8.6e28 
           & M Dwarf \\ \addlinespace

8          & UCAC3 53-724 & 43.9 \pm 0.2     
           & 2.3e28 & 5.8e28 & 1.1e29 
           & M Dwarf  \\
\bottomrule
\end{tabular*}
\caption{Assumed associations of stellar flares and physical properties of events: parallax-based distance, isotropic mm-wave luminosity $\nu L_\nu$, and type of star. 
All sources with types showing an asterisk have 2RXS X-ray sources within \ang{;1;}. Distances were pulled from the Gaia DR2 \citep{gaiadr2_18}.}
\label{table:properties}
\end{table*}

\begin{figure*}
  \centering
  \includegraphics[width=\linewidth]{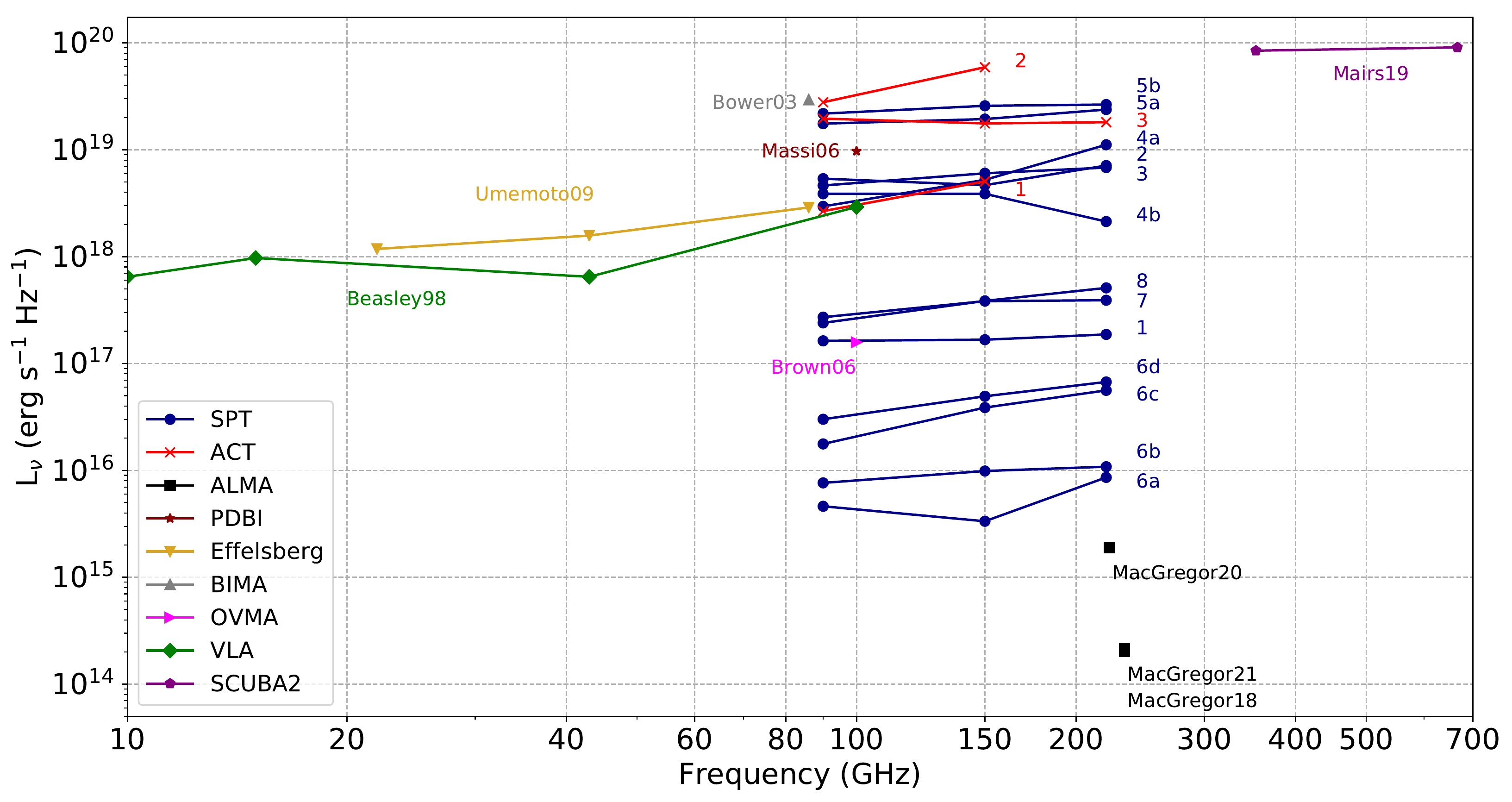}
  \caption{
    Luminosities per unit frequency $L_\nu$ of SPT stellar flares (\textit{blue}) and mm-wave stellar flares from the literature (\textit{remaining colors}) as a function of observing frequency. 
    Events are organized by instrument. SPT flares: from this work. ACT flares: \citet{naess2021}. ALMA: \citet{macgregor18,macgregor20,macgregor21}. PDBI: \citet{massi06}. Effelsberg: \citet{umemoto09}. BIMA: \citet{bower03}. OVMA: \citet{brown&brown06}. VLA: \citet{beasley&bastian98}. SCUBA-2: \citet{mairs19}.
    \label{fig:Lnu}}
\end{figure*}

The isotropic mm-wave luminosities $\nu L_\nu$ of the flaring events are shown in Table~\ref{table:properties} and range from roughly \SIrange{2e27}{6e30}{\erg \per \s} in the SPT-3G bands. 
At the bright end, this is comparable to previous mm-wave flares seen in RS CVn stars \citep{brown&brown06,beasley&bastian98} or T Tauri stars \citep{massi06,salter10,bower03}, although not as luminous as the sub-mm flare event in JW 566 \citep{mairs19}. 
The faint flares are brighter than those previously seen at similar wavelengths in M dwarfs \citep{macgregor20}. 
jhe isotropic luminosities per unit frequency $L_\nu$ of SPT stellar flares and selected mm-wave flares from the literature are compared in Figure~\ref{fig:Lnu}.

While the stars have a wide range in properties, there are some themes that emerge.
BI Ind (Source 2) is known to be of the type RS CVn and it has been suggested that both CX Ind (Source 3) and CD-55 8799 (Source 4) are RS CVn stars \citep{berdnikov2008}; this classification would be consistent with the similar energetics observed in the flares. 
BI Ind and the two historical RS CVn flare stars mentioned above are all in the giant branch of the HR diagram; however the possible RS CVn stars CX Ind and CD-55 8799 are redder and lower-luminosity than the typical giant.
The flare energy for BI Ind is comparable to the two historical flare stars. 
Two stars associated with mm-wave flares detected by ACT are also in that part of the HR diagram, with similar flare energy, although these stars have not previously been identified as RS CVn. 

Two sources (CX Ind, Source 3  and CC Eri, Source 6) are classified as BY Draconis-type variables in the SIMBAD\footnote{\url{http://simbad.u-strasbg.fr/simbad}} database, as is the previously known flare star AU Mic, which lies very close in the HR diagram to CC Eri. 
Two other sources are classified as ``Rotationally Variable,''  but are almost indistinguishable in the HR diagram from CX Ind. 
In terms of flare energy, CC Eri has substantially lower energy flux than the RS CVn stars, although more than ten times the flux of AU Mic, while CX Ind is energetically comparable to BI Ind. 
The remaining three stellar flares (Sources 1, 7, and 8) detected in this work are late M dwarfs, with one of them (UCAC4 114-133248, Source 1) actually a pair of M dwarfs.
The flare energy flux is comparable in these three cases (few $\times$ \SI[parse-numbers=false]{10^{28}}{\erg \per \second}), but in all cases at least a thousand times more luminous than the well-known Proxima Centauri mm-wave flare seen by \citet{macgregor18}.

\begin{figure}
\includegraphics[width=\linewidth]{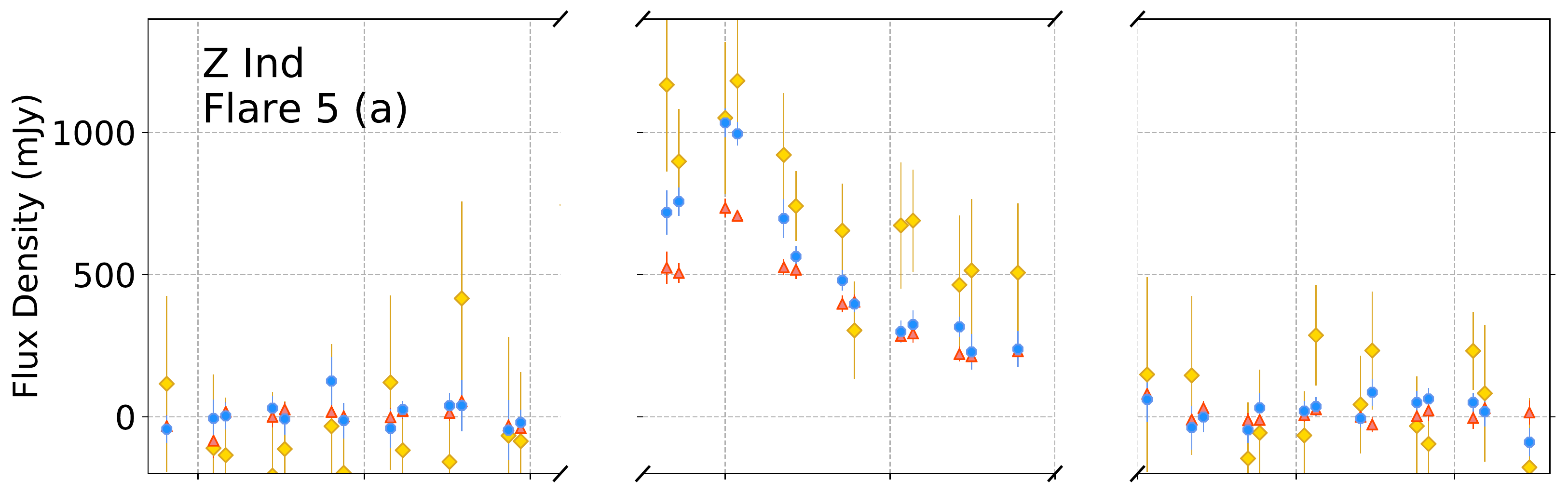}
\includegraphics[width=\linewidth]{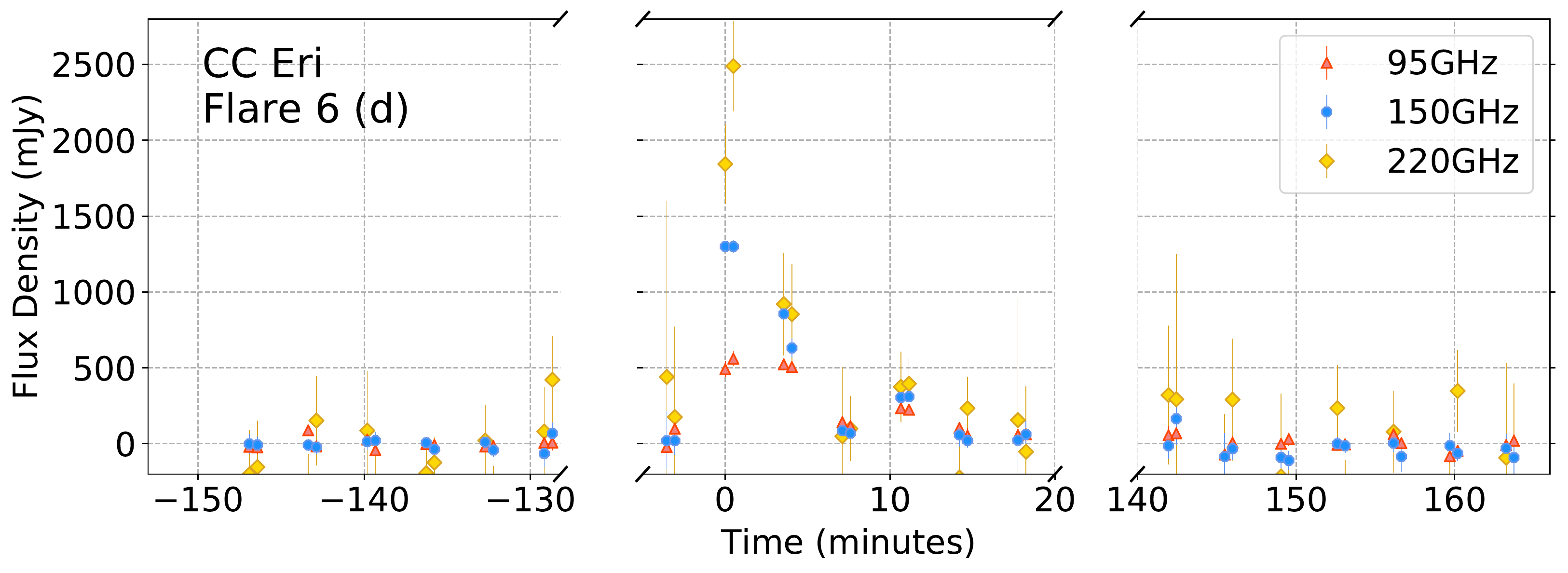}
\caption{Light curves for flare 5(a), associated with Z Ind, (top) and flare 6(d), associated with CC Eri, (bottom) showing flux densities derived from individual rasters over the source in 3 consecutive observations. 
The x-axis has been cut between observations, and shows the time in minutes since the flare peak at \SI{150}{GHz}, corresponding to MJD 59017.4011 for 5(a) and MJD 59158.6748 for 6(d).
\label{fig:singlescan_lightcurves}}
\end{figure}

As described in Section~\ref{sec:survey}, the SPT-3G raster scan observing strategy allows a limited ability to observe timescales shorter than {2} {hours} by examining the individual detector scans over a source position within an observation. 
There are typically 10 raster scans covering any given point in the subfield, which occur over $\sim$20 minutes of the 2-hour observation window.
Depending on the right ascension of the source, these rasters are spaced a maximum of {3} {minutes} apart. 
A preliminary examination of these data for flare 5(a), associated with Z~Ind, and flare 6(d), associated with CC~Eri, shows a true peak brightness exceeding \SI{1}{Jy} at \SI{150}{GHz}, with emission falling rapidly on 10-minute scales (see Figure~\ref{fig:singlescan_lightcurves}).
Due to the sub-observation timescales of some of these flares, the measured per-observation peak flux density, as reported in Table~\ref{tab:events}, is below the true peak amplitude. 
Future analyses may be able to trigger on such sub-observation data and provide a more detailed view of the sky at these minute scales than we present in this publication.

\begin{figure*}
\includegraphics[width=.5\linewidth]{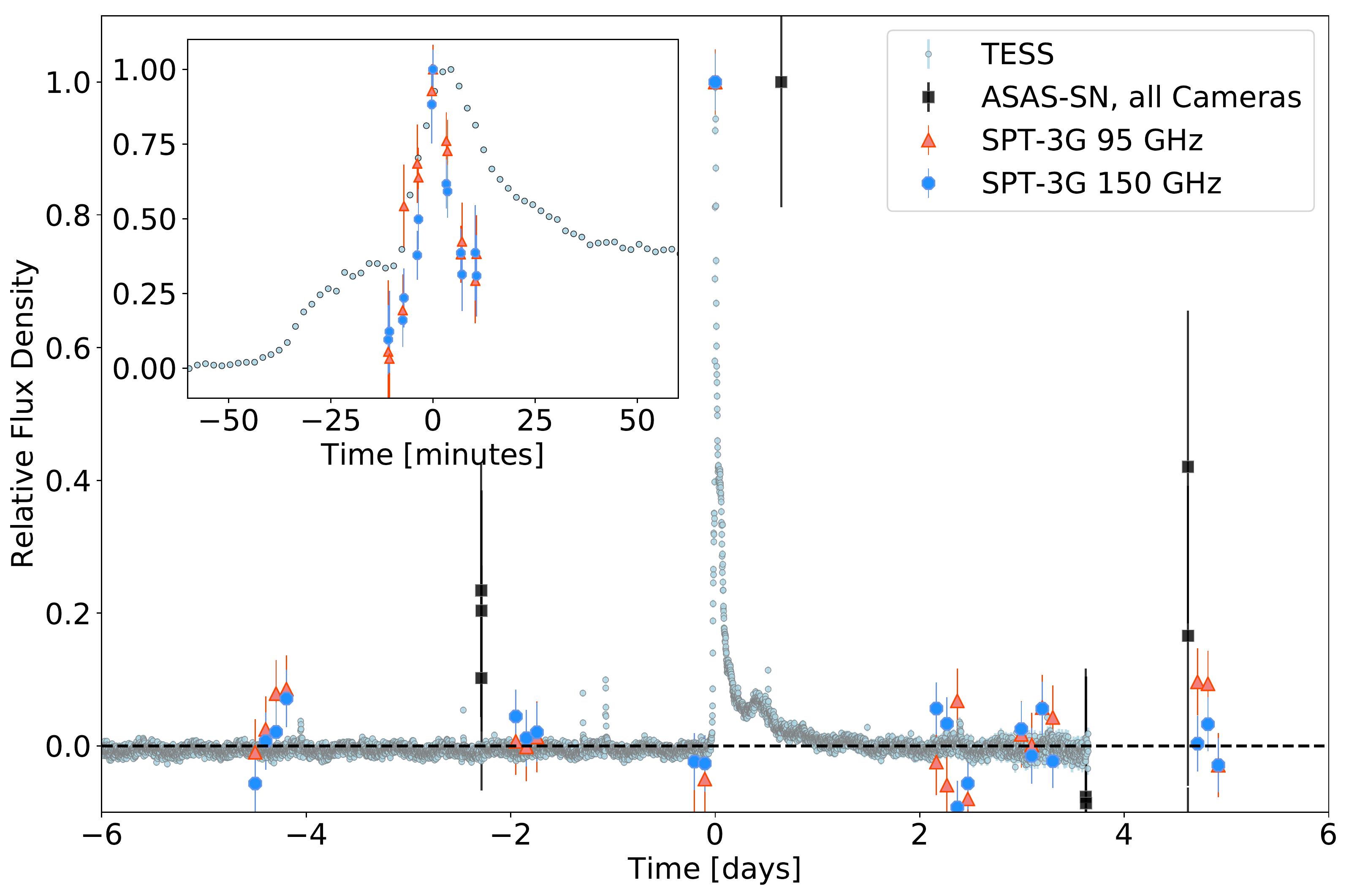}
\includegraphics[width=.5\linewidth]{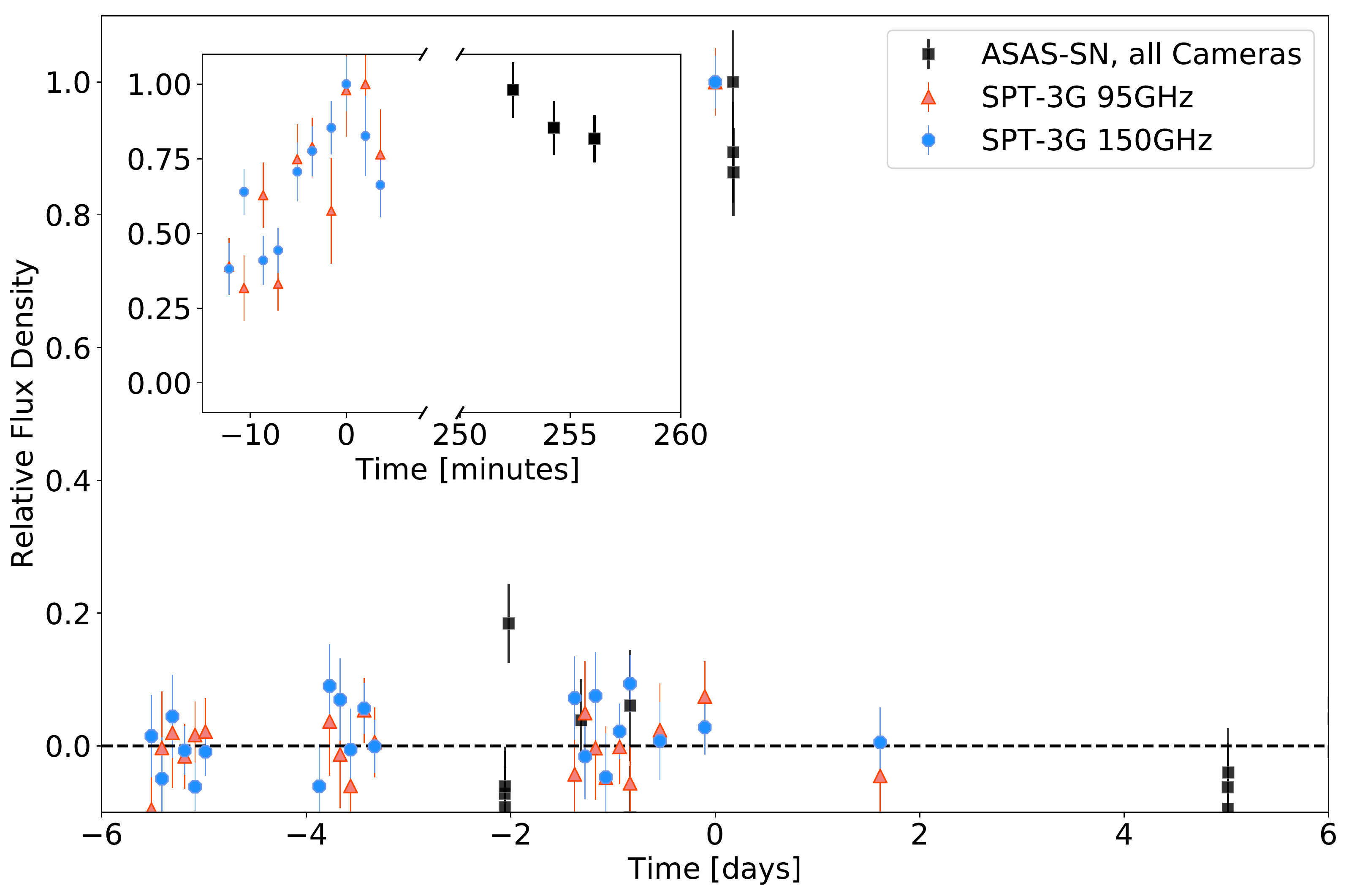}
\caption{Lightcurves for SPT-3G detected flares associated with WISE J025531.87-570252.3 (flare 7, left), and UCAC3 53-724 (flare 8, right). SPT-3G \SI{95}{GHz} (\textit{blue circles}) and \SI{150}{GHz} (\textit{red triangles}) data is plotted alongside ASAS-SN V-band data (\textit{black squares}) and, in the case of the WISE J025531.87-570252.3 associated flare, TESS V-band data (\textit{light blue circles}). In all cases, the flux density is mean-subtracted and plotted such that the maximum within \SI{\pm 2}{weeks} of the SPT-3G flare peak is normalized to 1. The x-axis shows time in days since the recorded flare peak at \SI{150}{GHz} (MJD 59110.2854 for flare 7, MJD 59167.0840 for flare 8), and the inset plot shows a zoomed-in region around the flare with single-scan SPT-3G flux density overplotted with the optical data. 
\label{fig:opticalcounterpartslc}}
\end{figure*}

We searched the ASAS-SN variable stars database for optical flux data near the peak times of the stellar flares. Six of the eight stars had some simultaneous ASAS-SN coverage. Of these, two show strong evidence for optical activity related to the millimeter-wave flares detected by SPT-3G: WISE J025531.87-5702523 (Source 7) and UCAC3 53-724 (Source 8), both M Dwarfs. Source 8 has a single ASAS-SN observation consisting of three successive 15-second exposures in V-band, showing a $5 \sigma$ increase in flux roughly 4 hours after the observed SPT-3G peak. There are large gaps in both the SPT-3G and ASAS-SN data, and no obvious conclusions can be drawn about the relation between the two detections. Source 7 has one ASAS-SN data point significantly above mean nearly 15 hours after the SPT-3G flare. Serendipitously, TESS had near-continuous coverage of Source 7 in the \SIrange{600}{1000}{nm} band for the entire duration of the SPT-3G observation, providing high-time-resolution data that shows a bright optical flare beginning some minutes before the first SPT-3G detection and decaying slowly over the next several hours \citep{brasseur19}. We show lightcurves for both of the flares with significant optical counterparts in Figure~\ref{fig:opticalcounterpartslc}. The data from different instruments are rescaled to allow comparison of the time behavior of both sources without making any inference about relative or absolute luminosities. As seen in the TESS data, the optical counterpart to the millimeter-wave flare of Source 7 starts rising an hour before the beginning of SPT-3G coverage. Starting at the first SPT-3G data point, both optical and mm rise rapidly until peaking 10 minutes later. The mm-wave lightcurve quickly falls back to half the observed peak flux before losing coverage, while the optical slowly decays and does not fall back to quiescence until some 24 hours after the peak.

\subsection{Extragalactic transients}
\label{sec:extragalactic}

\begin{figure}
\includegraphics[clip, trim = 2.5cm 0.35cm 2.5cm 1.1cm, width=\linewidth]{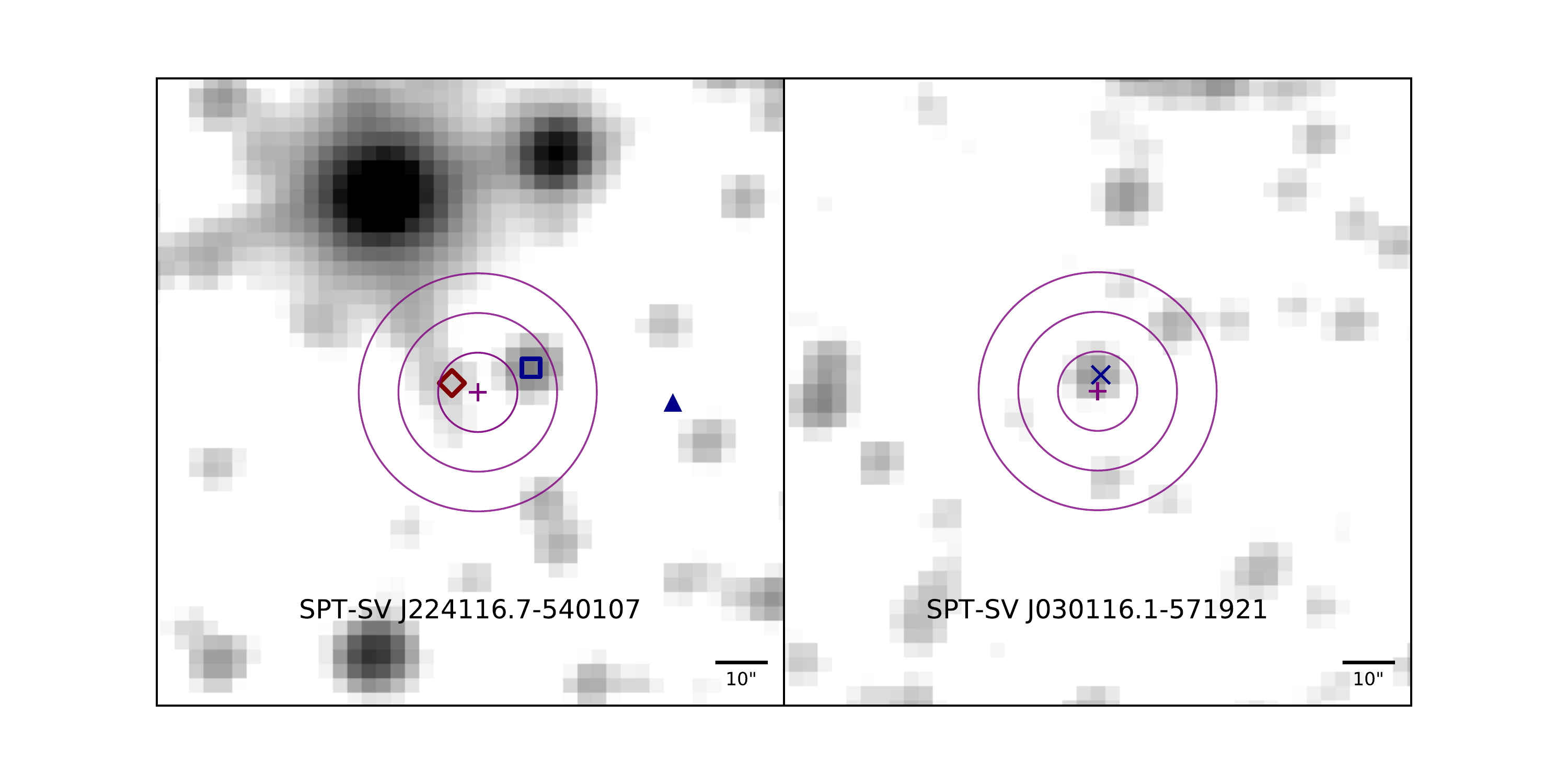}
\caption{
  Localization of the long-duration events for sources 9 (left) and 10 (right) using grayscale images from unWISE  \SI{3.4}{\mu m} W1 \citep{Lang_2014} in log stretch. The \textit{purple cross and contours} show the SPT-3G best-fit position and uncertainties in steps of $1\sigma$. 
  Positional uncertainties are derived from the test statistic map with an additional \ang{;;4.4} pointing uncertainty added in quadrature.
  For source 9 we overplot the positions of galaxies WISEA J224117.10-540105.2 (\textit{red diamond}) and WISEA J224115.38-540102.3 (\textit{blue square}), as well as the ROSAT X-ray source 2RXS J224112.8-540103 (\textit{blue triangle}) which has a positional uncertainty of \ang{;;\sim 21} and has been associated with the latter WISE galaxy.
  Source 10 is likely associated with the galaxy WISEA J030116.15-571917.7 (\textit{blue x}).
}
\label{fig:wise_panel_ldts}
\end{figure}

The two remaining sources are not obviously associated with any galactic source but, at low confidence, may be associated with WISE galaxies.
For these two longer-duration sources, the Fermi All-sky Variability Analysis \citep{abdollahi17} shows no significant associated gamma ray flare within several degrees of either source. In addition, no significant optical activity was seen in ASAS-SN for either source.
Source positions are determined from the peak of the likelihood surface ($\mathrm{TS}$ surface) created by applying the transient-finding algorithm to every pixel in a \ang{;3;}$\times$\ang{;3;} box around the source. Statistical positional uncertainties are expected to scale as the ratio of the beam width to signal-to-noise, and are estimated from the width of the $\mathrm{TS}$ surface using $\Delta \mathrm{TS}=2.3$ for a $\chi^2$ distribution with $2$ degrees of freedom.
To the extent that the flaring lightcurves approximate the Gaussian ansatz for the flare model, the $\mathrm{TS}$ map represents the optimally weighted combination of the different observing bands and periods in order to maximize localization precision.  
When we apply this method to the stars, where the association is unambiguous, there is an additional variance in position that is consistent with residual pointing uncertainty of \ang{;;4.4} in addition to the statistical uncertainties in localization. 
We add this additional uncertainty in quadrature to estimate the position uncertainties, finding sources 9 and 10 to have position uncertainties of \ang{;;7.6} and \ang{;;5.2}, respectively, as shown in Figure~\ref{fig:wise_panel_ldts}.

Source 9 (SPT-SV J224116.7-540107) is \ang{;;38} from a weak ROSAT X-ray emitter 2RXS J224112.8-540103, which has a positional uncertainty of \ang{;;\sim 21} \citep{2rxs2016}. 
This X-ray source has been associated with WISEA J224115.38-540102.3 \citep{salvato2018}, a galaxy that is \ang{;;12} from the SPT-3G position. 
There is another WISE galaxy that is closer, WISEA J224117.10-540105.2, at a separation of only \ang{;;4}, but 2 magnitudes fainter in WISE W1 (Band 1, \SI{3.4}{\mu m}). 
The larger sky density of such faint galaxies greatly increases the probability of chance alignment, even at this much closer distance. 
Using the local density of AllWISE sources \citep{cutri14} within a \ang{1} radius, the probability of a random AllWISE source being brighter and closer than either the dim or bright potential counterpart are \SI{6}{\percent} and \SI{8}{\percent}, respectively. 

Thus, it is not possible to make a definitive association of source~9 with a cataloged object.
Further study will be required to determine the counterpart for this source, for example with ALMA follow-up or detailed SED modeling using multiwavelength data.

Source 10 (SPT-SV J030116.1-57192) is within \ang{;;3} of the galaxy WISEA J030116.15-571917.7, with the next-closest source being \ang{;;17} away and substantially fainter (1.4 magnitudes in W1), making this association more secure.
The localizations of source 9 and 10 are shown in Figure~\ref{fig:wise_panel_ldts}.

The physical mechanism of the transient emission is unknown.
It is possible that the events are flares from AGN, which often have flaring behavior on this timescale.
A post-detection analysis revealed an average \SI{150}{GHz} flux density in 2019 of \SI{4.1 \pm 0.6}{mJy} and \SI{2.5 \pm 0.6}{mJy} for source 9 and 10 respectively.
The luminosity increase of source 9 from the 2019 average to the peak of the detected flare---a factor of 4---is at the upper limit of what is observed in brighter (\SI{> 10}{mJy}) sources monitored by SPT-3G.
Source 10 increased by a factor of 15, which is much larger than what is typical for bright AGN observed by the SPT or by \cite{trippe2011} in this band (even at the 95\% CL lower limit, which is still a factor of 7.4 increase) and may represent an origin different from ordinary AGN flaring or the potential for greater variability in faint AGN at millimeter wavelengths.

There is no cataloged radio source associated with either position, so both would require AGN with flat or rising spectra at radio wavelengths. 
Radio observations made with the 887~MHz Australian Square Kilometer Array Pathfinder (ASKAP) at points in time before and after the main mm flare of both sources (ASKAP observations on 28/29 March 2020 and 29/30 August 2020) do not show any evidence of either source at a depth of \SI{0.20}{mJy} \citep{hotan20}. 
No overlapping ASKAP observations were made during the peak period of June--July 2020.
Follow-up observations of these galaxies with deep radio and mm-wave observations may be able to identify possible AGN activity in these sources and shed light on whether the events observed here are part of some continuing flaring behavior from these objects.

The timescales and energies (assuming these sources are at an unremarkable redshift $z \lesssim 1$) are also consistent with expectations for tidal disruption events or an object like AT2018cow \citep{ho19}, but there were no transient alerts from observations at other wavelengths issued that match these objects.

\begin{figure}

\includegraphics[width=\linewidth]{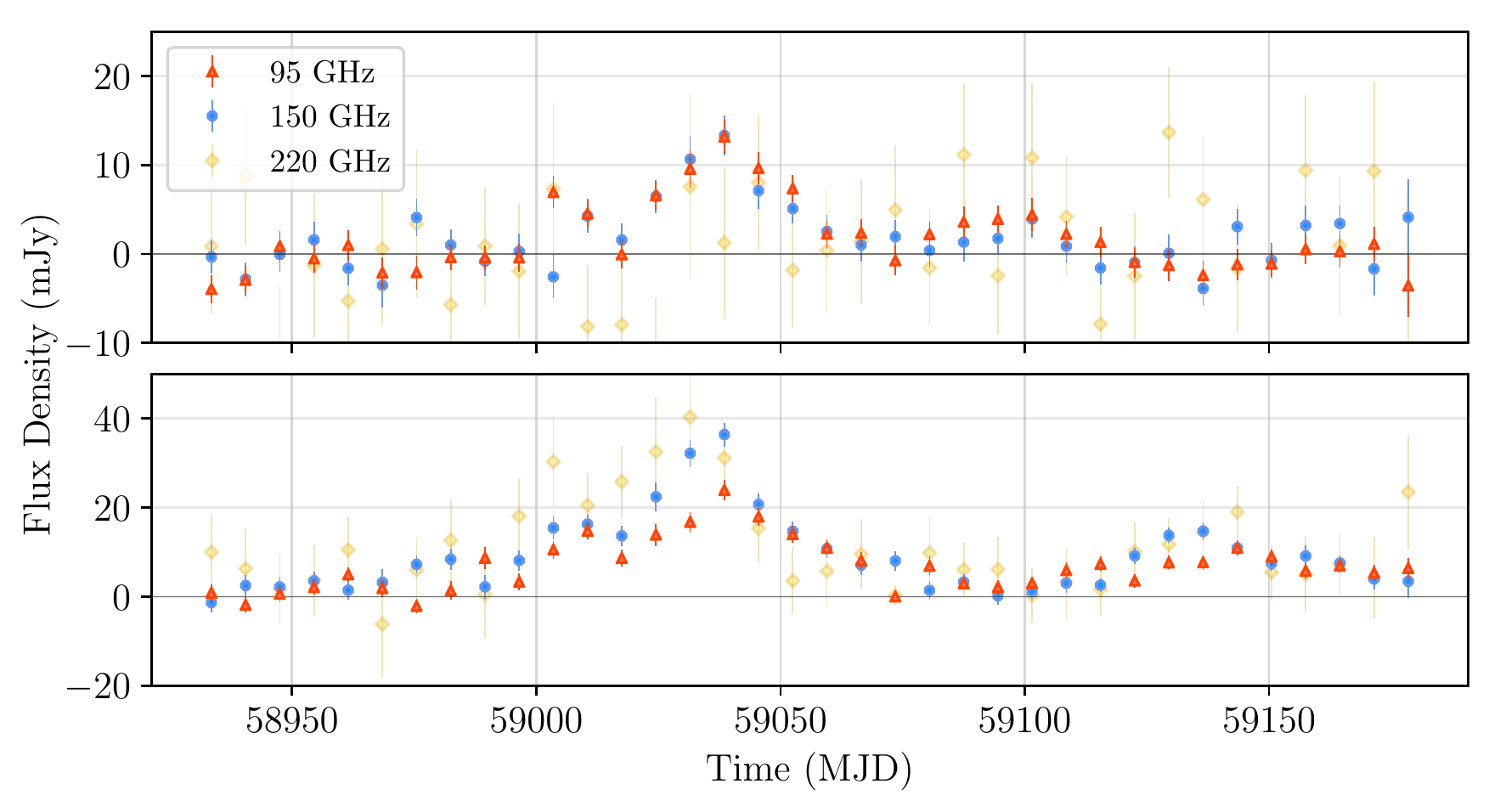}
\caption{Lightcurves of the SPT-3G transient events SPT-SV J224116.7-540107 (source 9, top) and SPT-SV J030016.1-571921 (source 10, bottom). 
\SI{95}{GHz} data are shown with \textit{red triangles}, \SI{150}{GHz} with \textit{blue circles}, and \SI{220}{GHz} with \textit{gold diamonds}. 
Each data point is a weighted average of all 2-hour field observations taken in a 7 day window centered at that time (x-axis) coordinate.}
\label{fig:source9n10}
\end{figure}

It is perhaps notable that both long-duration transient events, though \ang{35} apart on the sky, rise and fall with similar looking lightcurves and peak in the same week of 2020 (see Figure~\ref{fig:source9n10}). Given the 36-week observing period this is not an unlikely coincidence.
Additionally, most possible sources of systematic contamination can be eliminated by the fact that the two sources sit in different subfields. 
The center line of the SPT-3G footprint is at \ang{-56} declination, and the top and bottom half of the field use independent bolometer tunings, different HII calibration sources, and are observed on 
different observing days. The individual maps that contain the brightest transient observations show no signs of miscalibration, excess noise or excess pointing jitter, and other in-field sources have fluxes consistent with previous and subsequent observations. 
It is possible that some of the same physics is at play in these two sources to explain the similarity in flare shape and duration; further observations of similar events will provide more information.

\begin{figure}
\includegraphics[width=\linewidth]{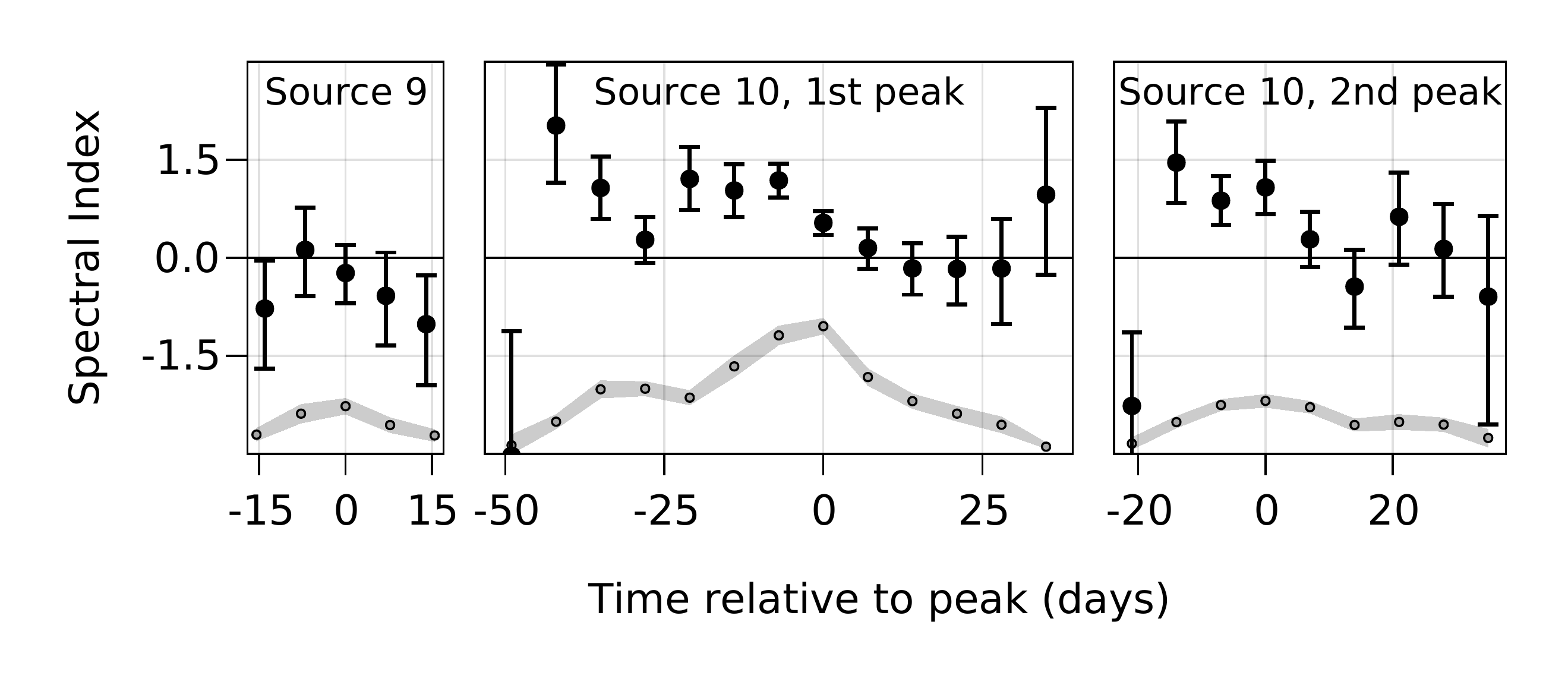}

\caption{
    Spectral index evolution of the two extragalactic transients.
    For each observation we plot the best-fit spectral index $\alpha$ (defined such that $\phi^b \propto b^\alpha$) in \textit{black} and the profile of the \SI{150}{GHz} flux density $\phi^{150}$ in \textit{grey} with error bars as described in Section~\ref{sec:methods}.
}
\label{fig:ldt_specinds}
\end{figure}

The emission spectrum of Source 10 shows a rising spectrum before the peak of the emission and a flat spectrum thereafter (Figure~\ref{fig:ldt_specinds}), while the dimmer Source 9 has large spectral uncertainties. 
This is consistent, though not uniquely so, with a self-absorbed synchrotron spectrum from a young cooling jet, with the initial brightening arising from falling self-absorption more than counteracting the cooling source and the peak of the spectrum moving through SPT-3G's observing bands at the peak emission time. 
No linear polarization was detected from the two sources at either the flare peak (Table~\ref{tab:specind_polfrac}) or when integrating over the flare, the latter approach giving $2 \sigma$ polarization fraction upper limits of 0.14, 0.12, and 0.49 at \SIlist{95;150;220}{GHz} respectively for Source 10, and 0.22 and 0.32 at \SIlist{95;150}{GHz} for Source 9.
These limits are weak enough that polarization information does not provide strong constraints on the emission mechanism or local magnetic field coherence in the emission region.

\section{Discussion and Conclusions}
\label{sec:conclusions}

The detection of 15 bright millimeter-wavelength flares in this work, many far above threshold for SPT-3G, suggests that these kinds of flares are common, and that a large number of sources remain to be detected. 
For example, a naive extrapolation of the rate of stellar flares seen in this paper---the rate of extragalactic transients is too small to draw a robust conclusion---would imply a rate of around a thousand flares of similar brightness per year over the whole sky. 
Further, the stellar flares seen here are short enough in time (minutes to hours) that many are missed by SPT-3G as a result of our observing cadence and analysis choices for this search. 
In addition to those missed because we are less sensitive to flares on timescales shorter than a few hours, our observing strategy has day-long gaps between re-observations of each subfield pair (Section~\ref{sec:survey}). 
This reduces the observing efficiency for hour-scale sources to approximately \SI{23}{\percent} over our \SI{1500}{\deg^2} survey, which implies that the true rate of bright stellar flares of the type seen here is at least 4000 per year on the full sky.

The extragalactic transients seen here are more of a puzzle.
The emission seen from both sources has a spectrum that evolves from rising to flat or falling, consistent with a newly emitted jet, and both sources show a second, smaller flare days to weeks afterwards.
Both sources were convincingly detected (though at low flux density) in the 2019 average map, but it is unclear what emission mechanism(s) caused the extreme flares seen here.
One possibility is that these flares were the result of regular AGN activity, but such large ratios of outburst to mean luminosity ($\sim 4$ and $\sim 15$ for sources 9 and 10, respectively) are rare:
Typical SPT-3G AGN fluxes vary by much smaller factors of $\lesssim 50\%$, with excursions to above 3 observed only in extreme cases (seen in $< 1\%$ of SPT-3G sources), and no sources seen with luminosity ratios above 4 when comparing 2020 peak to 2019 average flux data.
That sample, however, consists of brighter AGN and might not be representative of the unexplored population of faint AGN.

Such large luminosity variations are not unprecedented, especially over long timescales.
The transient source ACT-T J061647-402140\footnote{\url{https://www.astronomerstelegram.org/?read=12738}}, a possible mm-wave counterpart to the transient gamma-ray blazar Fermi 0617-4026, increased in brightness by a similar factor of $\num{\sim13}$ between June 2016 and January 2018.
Comparing the ACT flux densities with the 2010-2011 flux densities of the spatially coincident source SPT-S J061647-402147\footnote{\url{https://www.astronomerstelegram.org/?read=12837}} indicates an increase in flux by a factor of 15-20 over ${\sim}7$ years.
The emission seen is also too long in duration to be a GRB afterglow, which typically last for a few days \citep{ghirlanda13}.
Other possibilities, like a tidal disruption event, cannot be tested with the limited amount of data available. 

The SPT-3G camera will continue to observe this \SI{1500}{\deg^2} footprint until the completion of the survey at the end of 2023. 
This should at least quadruple the number of detected mm-wave transients with similar brightness, potentially probe new classes of variable mm-wave sources, and discover many more fainter sources as improvements in the analysis increase the sensitivity and time resolution of the search.
An already operating online alert system, using the methods described in this article, will soon provide public notice of these detections with latencies of \num{< 24} {hours}, enabling multi-wavelength follow-up to determine the nature of the emission seen in this work, as well as characterization of new sources while they are exhibiting variability.

\section*{Acknowledgments}
The authors thank Anna Ho for helpful comments on a draft version of this paper. We are also grateful to Jeff DeRosa and Johan Booth for providing guidance for South Pole weather balloons. Thanks to Charles Gammie, Leslie Looney, Paul Ricker, Bob Rutledge, and Laura Chomiuk for invaluable early discussions. 
The South Pole Telescope program is supported by the National Science Foundation (NSF) through grants PLR-1248097 and OPP-1852617, with this analysis and the online transient program supported by grant AST-1716965.
Partial support is also provided by the NSF Physics Frontier Center grant PHY-1125897 to the Kavli Institute of Cosmological Physics at the University of Chicago, the Kavli Foundation, and the Gordon and Betty Moore Foundation through grant GBMF\#947 to the University of Chicago.
Argonne National Laboratory's work was supported by the U.S. Department of Energy, Office of High Energy Physics, under contract DE-AC02-06CH11357.
Work at Fermi National Accelerator Laboratory, a DOE-OS, HEP User Facility managed by the Fermi Research Alliance, LLC, was supported under Contract No. DE-AC02-07CH11359.
The Cardiff authors acknowledge support from the UK Science and Technologies Facilities Council (STFC).
The IAP authors acknowledge support from the Centre National d'\'{E}tudes Spatiales (CNES).
JV acknowledges support from the Sloan Foundation.
The Melbourne authors acknowledge support from the Australian Research Council's Discovery Project scheme (DP200101068). 
The McGill authors acknowledge funding from the Natural Sciences and Engineering Research Council of Canada, Canadian Institute for Advanced Research, and the Fonds de recherche du Qu\'ebec Nature et technologies.
The UCLA and MSU authors acknowledge support from NSF AST-1716965 and CSSI-1835865.
This research was done using resources provided by the Open Science Grid \citep{pordes07, sfiligoi09}, which is supported by the NSF award 1148698, and the U.S. Department of Energy's Office of Science.
The data analysis pipeline also uses the scientific python stack \citep{hunter07, jones01, vanDerWalt11}.
This work has made use of data from the European Space Agency (ESA) mission
{\it Gaia} (\url{https://www.cosmos.esa.int/gaia}), processed by the {\it Gaia}
Data Processing and Analysis Consortium (DPAC,
\url{https://www.cosmos.esa.int/web/gaia/dpac/consortium}). 
Funding for the DPAC has been provided by national institutions, in particular the institutions participating in the {\it Gaia} Multilateral Agreement.
This publication makes use of data products from the Wide-field Infrared Survey Explorer, which is a joint project of the University of California, Los Angeles, and the Jet Propulsion Laboratory/California Institute of Technology, and NEOWISE, which is a project of the Jet Propulsion Laboratory/California Institute of Technology. 
WISE and NEOWISE are funded by the National Aeronautics and Space Administration.
This research has made use of the NASA/IPAC Infrared Science Archive, which is operated by the Jet Propulsion Laboratory, California Institute of Technology, under contract with the National Aeronautics and Space Administration.
This paper includes data collected by the TESS mission. Funding for the TESS mission is provided by the NASA's Science Mission Directorate.

\facilities{ASAS, ASKAP, Fermi, Gaia, NEOWISE, ROSAT, SPT (SPT-3G), TESS, WISE}

\end{document}